\documentclass[reprint,amsmath,amssymb,showkeys,showpacs]{revtex4-1}
\usepackage{graphicx}% Include figure files
\usepackage{float}
\usepackage{pxfonts}
\usepackage{stmaryrd}
\usepackage{mathptmx}
\begin{document}
\title{Generating multi-dimensional entangled states of atoms under large atom-cavity detuning}
\author{Peng Shi$^{1}$, Li-Bo Chen$^{2}$, Wen-Dong Li$^{1}$, Chun-Nian Ren$^{1}$, Chun-Hong Zheng$^{1}$}
\author{Yong-Jian Gu$^{1}$}
\email{E-mail address: yjgu@ouc.edu.cn}
\address{$^{1}$Department of Physics, Ocean University of China, Qingdao 266100, China\\$^{2}$School of Science, Qingdao Technological University, Qingdao 266033, China}

\date{September 1, 2012}

\begin{abstract}
We propose a scheme to deterministically generate two-dimensional and three-dimensional entangled states of atoms by passing two $^{87}Rb$ atoms through a high-$Q$ bi-mode cavity alternately. The atomic spontaneous decay is efficiently suppressed because of large atom-cavity detuning in our scheme. With the strictly numerical simulation, it shows that, although the cavity decay exists unavoidably, our proposal is good enough to demonstrate the generation of atomic entanglement with high fidelity and within the current experimental technologies.
\end{abstract}

\maketitle

\section{Introduction}
Quantum entanglement plays a vital role in many practical quantum information systems \cite{CHBennett,CHBennett2,AKEkert}. Cavity quantum electrodynamics ($QED$) provides an almost ideal system for the generation of entangled states \cite{ARauschenbeutel,MSZubairy,QATurchette}. Atoms trapped in an optical cavity, a basic model in cavity $QED$, are believed to be a promising system for quantum computation and quantum communication \cite{EKnill,CCGerry,TPellizzari,SHughes}, where entangling of atoms is a fundamental operation \cite{SBZheng}. In addition to the entangling of two-dimensional (qubit) atoms \cite{FFrancica}, the entangling of higher-dimensional atoms also attracts a lot of attention \cite{ADelgado,XBZou,SBZheng2,SBZheng3,XMLin,GWLin,SYYe}. These higher-dimensional entanglement systems, such as three-dimensional (qutrit) entangle states of atoms, are of great interests owing to the extended possibilities they provide, which including higher information density coding \cite{TDurt}, stronger violations of local realism \cite{DKaszlikowski,DCollins}, and more resilience to error than two-dimensional systems \cite{MFujiwara}.  Recently some theoretical schemes have been proposed for implementing three-dimensional entangled state of atoms in a cavity $QED$ system via adiabatic passage \cite{LBChen,LBChen2}, or with spin qubits coupled to a bimodal microsphere cavity \cite{ASZheng}. However, the main problems for these entangling atoms in cavities are the unavoidable decoherence \cite{HMabuchi} due to the leakage of photons from the cavity modes, or the imperfect suppression of the spontaneous radiation in the atoms. 

In order to suppress these disadvantages, we use the adiabatical state evolution under large detuning to generate atomic entangled states. One of the distinct advantages of our proposal is that the excited states can be effectively eliminated, thus the atomic spontaneous emission does not play an important role; moreover, in our scheme we implement not only a two-dimensional entanglement but also a three-dimensional entanglement of two $^{87}Rb$ atoms, and our computer numerical simulation results indicate that by choosing proper parameters we could deterministically entangle two atoms with muliti-dimensions in an optical cavity with high fidelities, although the cavity decay exists unavoidably; most of all, the experimental requirements mentioned in our scheme are approximately feasible at present and therefore our scheme should be realizable in the near future.

\section{The fundamental model}

The schematic represention is shown in Fig.1, where two atoms A and B fall through the empty bi-mode optical cavity C alternately. The relevant atomic levels and transitions are also depicted in this figure, such level structures can be achieved in $^{87}Rb $ \cite{TWilk,TWilk2,BWeber}. The states $|g_L\rangle$, $|g_0\rangle$, $|g_R\rangle$ and $|g_a\rangle$ correspond
to $^{87}Rb$ atom hyperfine levels $|F=1,m_F=-1\rangle$, $|F=1,m_F=0\rangle$, $|F=1,m_F=1\rangle$ of $5S_{1/2}$ and $|F=2,m_F=0\rangle$ of $5S_{1/2}$, while $|e_L\rangle$, $|e_0\rangle$ and $|e_R\rangle$ correspond to $|F=1,m_F=-1\rangle$, $|F=1,m_F=0\rangle$ and $|F=1,m_F=1\rangle$ of $5P_{3/2}$.
\begin{figure}[H]
\centering
\includegraphics[height=5.5cm, width=8cm]{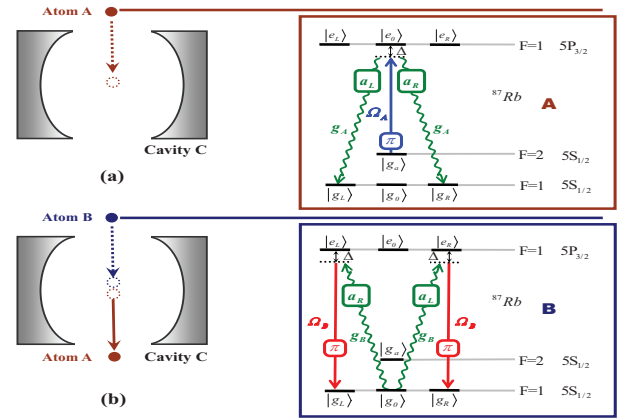}
\caption{(Color online) Schematic illustration of generating atomic multi-dimensional entangled states with two $^{87}Rb$ atoms in a bi-mode cavity. (a) The atom A enters the cavity C, and interacts with a $\pi$-polarized classical pump field. (b) After the atom A passes through the cavity C, the atom B enters the cavity and interacts with the cavity mode. The atomic levels and transitions of the atoms involved are shown as well, $\Delta$ is the detuning between the cavity mode and the corresponding atomic transition.}
\label{fig:}
\end{figure}
Initially, the cavity mode is in the vacuum state, and the atoms A and B are prepared in the state $|g_a\rangle_A$ and $|g_0\rangle_B$ respectively before they fall into the empty cavities C. The following atoms passing process can be divided into two stages.

In the first stage, as shown in Fig.1 (a), the atom A enters the cavity C first, the atomic transition $|g_a\rangle_A$ $\leftrightarrow$ $|e_0\rangle_A$ is driven resonantly by a $\pi$-polarized (line polarized) classical field with Rabi frequency $\Omega_A$; the atomic transitions $|e_0\rangle_A$ $\leftrightarrow$ $|g_R\rangle_A$ and $|e_0\rangle_A$ $\leftrightarrow$ $|g_L\rangle_A$ are coupled to the right and left circular polarized cavity mode with the same conherent coupling constant $g_A$, where we assume $\Omega_A$ and $g_A$ are all real \cite{JShu}. The Hamiltonian for the system in the interaction picture can be written as (here we set $\hbar=1$)
\begin{equation}
\begin{split}
H^A_I = &\Omega_A|e_0\rangle_A\langle g_a|e^{-i\Delta t}+g_A a_R |e_0\rangle_A\langle g_R|e^{-i\Delta t}+ g_A a_L \\&{\times} |e_0\rangle_A\langle g_L|e^{-i\Delta t}+H.c.,
\end{split}
\end{equation}
where $a_R$, $a_L$ are the annihilation operators of the two opposite right, left circular polarizations respectively; $\Delta$ is the detuning between the cavity mode and the corresponding atomic transition. In the case of large atom-cavity detuning, i.e., $\Delta\gg g_A$, the excited states can be eliminated adiabatically to obtain the effective Hamiltonian
\begin{equation}\label{formula1}
\begin{split}
H^A_{eff}=&\frac{\Omega_A^2}{\Delta}|g_a\rangle_A\langle g_a|+\frac{g_A^2}{\Delta}(a_L^{\dag} a_L |g_L\rangle_A\langle g_L|+a_R^{\dag} a_R |g_R\rangle_A\\&{\times} _A\langle g_R|+a_L^{\dag} a_R |g_L\rangle_A\langle g_R|+a_R^{\dag} a_L |g_R\rangle_A\langle g_L|)\\& {+}\frac{\Omega_A g_A}{\Delta}(a_R^{\dag} |g_R\rangle_A\langle g_a|+a_L^{\dag} |g_L\rangle_A\langle g_a|+H.c.).
\end{split}
\end{equation}

Then we switch to the second stage, as shown in Fig.1 (b), the atom A passes through the cavity C, meanwhile, the atom B enters the cavity C and interacts with the cavity mode. The atomic transitions $|g_0\rangle_B$ $\leftrightarrow$ $|e_L\rangle_B$ and $|g_0\rangle_B$ $\leftrightarrow$ $|e_R\rangle_B$ are resonantly coupled to the right and left circular polarized cavity mode with the same conherent coupling constant $g_B$; the atomic transitions $|e_L\rangle_B$ $\leftrightarrow$ $|g_L\rangle_B$ and $|e_R\rangle_B$ $\leftrightarrow$ $|g_R\rangle_B$ are driven resonantly by a $\pi$-polarized (line polarized) classical field with Rabi frequency $\Omega_B$, where we also assume $\Omega_B$ and $g_B$ are all real. The Hamiltonian for the system of second stage in the interaction picture is (set $\hbar=1$)
\begin{equation}
\begin{split}
H^B_I=&\Omega_B(|e_L\rangle_B\langle g_L|e^{-i\Delta t}+|e_R\rangle_B\langle g_R|e^{-i\Delta t})+g_B a_L |e_R\rangle_B\\&{\times} _B\langle g_0|e^{-i\Delta t}+g_B a_R |e_L\rangle_B\langle g_0|e^{-i\Delta t}+H.c.,
\end{split}
\end{equation}
similarly, in the case of $\Delta\gg g_B$, the effective Hamiltonian of the second stage system can be easily obtained
\begin{equation}\label{formula2}
\begin{split}
H^B_{eff}=&\frac{\Omega_B^2}{\Delta}(|g_L\rangle_B\langle g_L|+|g_R\rangle_B\langle g_R|)+\frac{g_B^2}{\Delta}(a_R^{\dag} a_R |g_0\rangle_B\\&{\times} _B\langle g_0|{+}a_L^{\dag} a_L |g_0\rangle_B\langle g_0|)+\frac{\Omega_B g_B}{\Delta}(a_R|g_L\rangle_B\langle g_0|\\&+ a_L|g_R\rangle_B\langle g_0|+H.c.).
\end{split}
\end{equation}

\section{Generating a two-dimensional entangled state of atoms}\label{section2}
According to the descriptions above, at time $t=0$, the initial state of the system is $|\psi(0)\rangle$ = $|g_a\rangle_A$ ${\varotimes}$ $|0\rangle_C$, where $|0\rangle_C$ is the vacuum state of the cavity C. At the end of the first stage, governed by $H^A_{eff}$ described in Eq.{\eqref{formula1}}, the system evolves to the state $
|\psi(t)\rangle$ = $c_1(t)|g_a\rangle_A|0\rangle_C$ + $c_2(t)|g_L\rangle_A|L\rangle_C$ + $c_3(t)|g_R\rangle_A|R\rangle_C$, where $|L\rangle_C$ and $|R\rangle_C$ are respectively the left and right circular cavity modes. The indeterminate coefficients $c_1(t)$ = $(2g_A^2+\Omega_A^2 e^{-i\eta t{/}\Delta})/{\eta}$, $c_2(t)$ = $c_3(t)$ = $(g_A {\Omega}_A e^{-i\eta t{/}{\Delta}}-g_A \Omega_A)/{\eta}$, where we take $\eta=2g_A^2+{\Omega}_A^2$ for convenience. If we choose $\Omega_A=\sqrt{2} g_A$, at time $t_1={\pi} {\Delta}/{4g_A^2}$, the system initial state $|{\psi}(0)\rangle$ transfer to a superposition
\begin{equation}
|\psi(t_1)\rangle=\frac{1}{\sqrt{2}}(|g_L\rangle_A|L\rangle_C+|g_R\rangle_A|R\rangle_C),
\end{equation}
which is the entangled state of the atom A and the cavity modes, where we have neglected the common factor -1.

Nest stage, the atom A passes through the cavity, instead, the atom B falls into the cavity. It is obvious that the state of the system can be expressed as $|\psi(t_1')\rangle$ = $1/{\sqrt{2}}$ ($|g_L\rangle_A|L\rangle_C$ + $|g_R\rangle_A|R\rangle_C$) ${\varotimes}$ $|g_0\rangle_B$ for the present, which is a product state of the entangled state $|\psi(t_1)\rangle$ and the ground state $|g_0\rangle_B$ of the atom B, where $t_1'> t_1$ owing to time delay before the atom B enters the cavity. Governed by $H^B_{eff}$ described in Eq.{\eqref{formula2}}, the whole system evolves to the state
$|\psi(t)\rangle$ = $d_1(t)|g_L\rangle_A|g_0\rangle_B|L\rangle_C$ + $d_2(t)|g_R\rangle_A|g_0\rangle_B|R\rangle_C$ + $d_3(t)|g_L\rangle_A|g_R\rangle_B|0\rangle_C$ + $d_4(t)|g_R\rangle_A|g_L\rangle_B|0\rangle_C$, where the coefficients $d_1(t)$ = $d_2(t)$ = ($\Omega_B^2$ + $g_B^2$ $e^{-i\xi t{/}\Delta}$)/${\sqrt{2}\xi}$, $d_3(t)$ = $d_4(t)$ = ($g_B \Omega_B$ $e^{-i\xi t{/}\Delta}$ - $g_B \Omega_B$)/${\sqrt{2}\xi}$, and we take $\xi=g_B^2+\Omega_B^2$ for convenience. If we choose $\Omega_B = g_B$, at time $t_2=t_1'+{\pi} {\Delta} {/} {2g_B^2}$, we can obtain a two-qubit atomic entangled state
\begin{equation}
|\psi(t_2)\rangle=\frac{1}{\sqrt{2}}(|g_L\rangle_A|g_R\rangle_B+|g_R\rangle_A|g_L\rangle_B){\varotimes}|0\rangle_C,
\end{equation}
where we have neglected the common factor -1.

\begin{figure}[H]
\centering
\includegraphics[width=8.5cm]{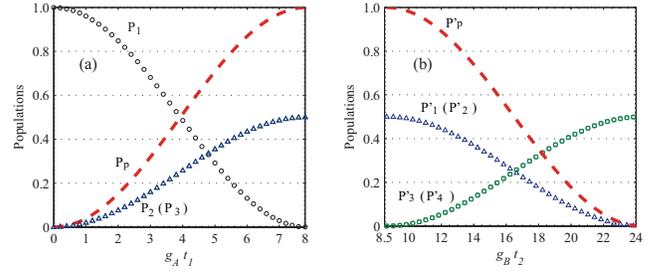}
\caption{(Color online) Time-dependent atomic-photonic-state populations vs dimensionless time $g_At_1 (g_Bt_2)$ in the first stage (a) and the second stage (b), where $\Omega_A=\sqrt{2}g_A$, $\Omega_B=g_B$, $\Delta$ = $10g_A$ = $10g_B$, and $t_1'- t_1$ ${\approx}$ $0.1t_1$. $P_{i} (i=1,2,3)$ and $P_{j}' (j=1,2,3,4,)$ respectively denote the populations of the states $|g_a\rangle_A|0\rangle_C$, $|g_L\rangle_A|L\rangle_C$, $|g_R\rangle_A|R\rangle_C$, $|g_L\rangle_A|g_0\rangle_B|L\rangle_C$, $|g_R\rangle_A|g_0\rangle_B|R\rangle_C$, $|g_L\rangle_A|g_R\rangle_B|0\rangle_C$ and $|g_R\rangle_A|g_L\rangle_B|0\rangle_C$. $P_p$ and $P_p'$ are the probabilities with which one photon appears in the cavity C related to the two stages.}
\label{fig:}
\end{figure}

Fig.2 shows the numerical simulation results of the two-qubit entanglement generation process. Fig.2 (a) shows the time evolution of populations in the first stage. Time fulfilling $g_At_1=5\pi/2$, then $P_1$ approaches 0, $P_2$ and $P_3$ arrive at the maximum value 0.5 simultaneously, at this time the system emits a photon. Fig.2 (b) shows the time evolution of the populations in the second stage, at time $g_Bt_2=5\pi+g_Bt_1'$, $P_1'$ and $P_2'$ approaches 0, while $P_3$ and $P_4$ arrive at the maximum value 0.5 simultaneously, that is to say, the state of the system transfers to the superposition of states $|g_L\rangle_A|g_R\rangle_B|0\rangle_C$ and $|g_R\rangle_A|g_L\rangle_B|0\rangle_C$ with the same probability of 1/2, which means the successful generation of the two-dimensional entangled state, the photon in cavity disappears in this process.

\section{Generating a three-dimensional entangled state of atoms}
In this section, we will consider another kind of circumstance. Return to the first stage in section \ref{section2}, differently from what we have done before, if we choose $\Omega_A=(1+\sqrt{3}) g_A$, change time $t_1={\pi} {\Delta} {/} {2(3+\sqrt{3}) g_A^2}$, then the final state of the system in the first stage can be rewritten as
\begin{equation}
|{\psi}'(t_1)\rangle=\frac{1}{\sqrt{3}}(|g_a\rangle_A|0\rangle_C+|g_L\rangle_A|L\rangle_C+|g_R\rangle_A|R\rangle_C.
\end{equation}

Nest stage, Similarly governed by $H^B_{eff}$ the whole system evolves to the new state $|{\psi}'(t){\rangle}$ = ${d_1}'(t)$ ${|g_a\rangle_A}{|g_0\rangle_B}{|0\rangle_C}$ + ${d_2}'(t)$ ${|g_L\rangle_A}{|g_0\rangle_B}{|L\rangle_C}$ + ${d_3}'(t)$ ${|g_R\rangle_A}{|g_0\rangle_B}{|R\rangle_C}$ + ${d_4}'(t)$ ${\times}$${|g_L\rangle_A}{|g_R\rangle_B}{|0\rangle_C}$ + ${d_5}'(t)$ ${|g_R\rangle_A}{|g_L\rangle_B}{|0\rangle_C}$ from changed initial state $|{\psi}'(t_1')\rangle$ = $|{\psi}'(t_1)\rangle$ ${\varotimes}$ $|g_0\rangle_B$ of the second stage, where  $t_1'$ $>$ $t_1$ is the initial time of this stage like section \ref{section2}. The indeterminate coefficients ${d_1}'(t)$ = ${1}/{\sqrt{3}}$, ${d_2}'(t)$ = ${d_3}'(t)$ = (${\Omega}_B^2$ + $g_B^2 e^{-i\zeta t{/}{\Delta}}$)/${\sqrt{3}\zeta}$, ${d_4}'(t)$ = ${d_5}'(t)$ = ($g_B {\Omega}_B$ $e^{-i\zeta t{/}\Delta}$ - $g_B {\Omega}_B$)/${\sqrt{3}\zeta}$, if we set $\zeta=g_B^2+\Omega_B^2$. As the same value as we choose $\Omega_B = g_B$ in section \ref{section2}, at time $t_2$ = $t_1'$ + ${\pi} {\Delta} {/} {2g_B^2}$, we can also obtain a superposition state

\begin{equation}
\begin{split}
|{\psi}'(t_2)\rangle=&\frac{1}{\sqrt{3}}(|g_a\rangle_A|g_0\rangle_B+|g_L\rangle_A|g_R\rangle_B+|g_R\rangle_A|g_L\rangle_B)\\&{\varotimes}|0\rangle_C,
\end{split}
\end{equation}
which is a product state of a two-qutrit atomic entangled state and the cavity mode vacuum state.
\begin{figure}[H]
\centering
\includegraphics[width=8.5cm]{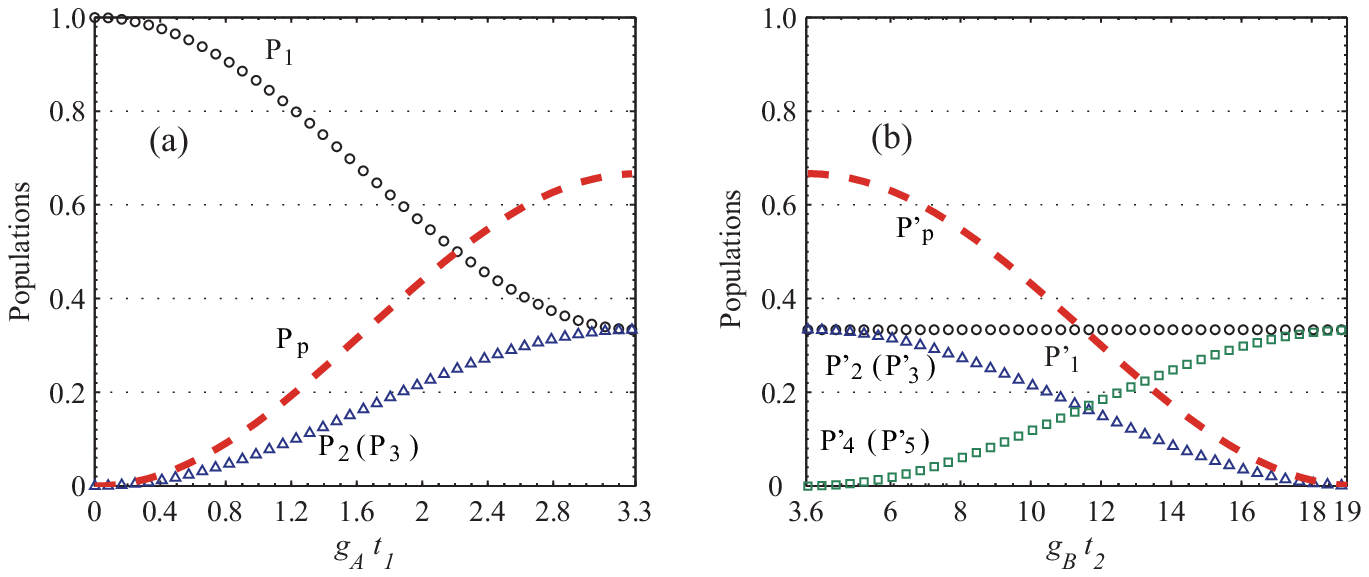}
\caption{(Color online) Time-dependent atomic-photonic-state populations vs dimensionless time $g_At_1$ $(g_Bt_2)$ in the first stage (a) and the second stage (b), where $\Omega_A$ = $(1+\sqrt{3})g_A$, $\Omega_B=g_B$, ${\Delta}$ = ${10g_A}$ = ${10g_B}$, and $t_1'-t_1$ ${\approx}$ $0.1 t_1$. $P_i$ ${(i=1,2,3)}$ and $P'_j$ ${(j=1,2,3,4,5)}$ respectively denote the populations of states $|g_a\rangle_A|0\rangle_C$, $|g_L\rangle_A|L\rangle_C$, $|g_R\rangle_A|R\rangle_C$, $|g_a\rangle_A|g_0\rangle_B|0\rangle_C$, $|g_L\rangle_A|g_0\rangle_B|L\rangle_C$, $|g_R\rangle_A|g_0\rangle_B|R\rangle_C$, $|g_L\rangle_A|g_R\rangle_B|0\rangle_C$, and $|g_R\rangle_A|g_L\rangle_B|0\rangle_C$. $P_p$ and $P_p'$ are the probabilities with which one photon appears in the cavity C related to the two stages.}
\label{fig:}
\end{figure}
Fig.3 shows the numerical simulation results of the two-qutrit entanglement generation process. In the first stage, as is shown in Fig.3 (a), time fulfilling $g_At_1=5\pi/(3+\sqrt{3})$, then $P_1$, $P_2$ and $P_3$ approaches 1/3 simultaneously, and the photon appears in the cavity C with the probability of 2/3. In the second stage, Fig.3 (b) shows that at time $g_Bt_2=5\pi+g_Bt_1'$, $P_2'$ and $P_3'$ approaches 0, while $P_1'$, $P_4'$ and $P_5'$ approaches 1/3 simultaneously (actually, $P_1'$ always equals 1/3 in this stage), that is to say, the state of the system transfers to the superposition state $1/\sqrt{3}$ ($|g_a\rangle_A|g_0\rangle_B$ + $|g_L\rangle_A|g_R\rangle_B$ + $|g_R\rangle_A|g_L\rangle_B$) ${\varotimes}$ $|0\rangle_C$, which means the successful generation of the three-dimensional entangled state, and the cavity mode returns back to initial vacuum state.

\section{Discussion and conclusion}
To evaluate the performance of our scheme, we now give an indispensable discussion on the dissipative processes in the whole interaction time: spontaneous atomic decay from excited states at the common rate $\gamma$ and cavity decay rate $\kappa$. As we use the large detuning approximation, the atomic spontaneous emission can be neglected ($\gamma \approx 0$). The evolution of the system can be described by a non-hermitian conditional Hamiltonian \cite{MBPlenio,JCho}, which is be simplified as $H_{con}^{A(B)}$=$H_{eff}^{A(B)}$ $-$ $i{\kappa}$ (${a_L}^{\dag}a_L$ + ${a_R}^{\dag}a_R$). 

Starting with the initial state $|\psi(0)\rangle$, under the conditional hamiltonian, the system evolves at the time $t$ to $|\psi(t)\rangle$=$e^{-i H_{con}^{A(B)}t}$ $|\psi(0)\rangle$. The success probability and the fidelity of the whole process are given by $P$=$|\langle\psi(t)|\psi(t)\rangle|^2$ and $F$ =$|\langle\psi_{id}(t)|\psi(t)\rangle|^2$, where $\psi_{id}(t)$ is the ideal state. Fig.4 shows the success possibility and the fidelity of the two-qubit (qutrit) entanglement in our scheme as a function of the cavity decay rate $\kappa$.

It is necessary to discuss the effect of atomic motion and the experimental feasibility of our scheme. For one thing, trapping atoms in cavity $QED$ has been realized in early experiments \cite{JVolz,PMaunz,MKhudaverdyan,MKhudaverdyan2}. Our scheme needs atoms to pass through one cavity separately, in current optical cavity $QED$ systems, the coupling strength $g(\vec{r})$ depends on the atomic position $\vec{r}$. If atomic motion is localized to a size $\Delta \vec{r}$ comparable to the cavity mode resonant wavelength ${\lambda}_c$, this position-dependent uncertainty generally spoils the quantum coherence or entanglement. To enforce an approximate coupling constant $g$, one requires the Lamb-Dicke limit $\Delta \vec{r}$$\ll$${\lambda}_c$ to suppress the decoherence caused by atomic motions. Actually, this requirement has been met in a one-cavity system by the current experimental conditions, and can be obtained by atom-cooling techniques. For another, from the former numerical simulation, we need $\kappa$ smaller than $g$ by approximately two orders of magnitude. Based on the experiments about realizing high-Q cavity and strong atom-cavity coupling \cite{SMSpillane,JRBuck}, the condition $(g,\kappa,\gamma)/2\pi= (750,2.6,3.5) MHz$ is realizable. Under such condition, we can entangle the atoms with the fidelity larger than 99.2\% in our scheme. 

It should be noticed that although we have neglected the effect of atomic spontaneous emission due to large detuning, the atom-cavity system is still in the strongly cooperative limit
($g^2/{\kappa}{\gamma}{\gg}1$). We can adiabatically eliminate the excited states \cite{BSun,JShu}, and obtain effective coupling strength ${\Omega}_{eff}={\Omega}g/{\Delta}$ and effective atomic decay rate ${\gamma}_{eff}={\Omega}^2{\gamma}/{\Delta}^2$. To deterministically generate an entangled state, one requires the strong coupling conditions ${\Omega}_{eff}{\gg}{\gamma}_{eff}$ and ${\Omega}_{eff}{\gg}{\kappa}$, which reduce to the strongly cooperative limit.
\begin{figure}[t]
\centering
\includegraphics[width=8.5cm]{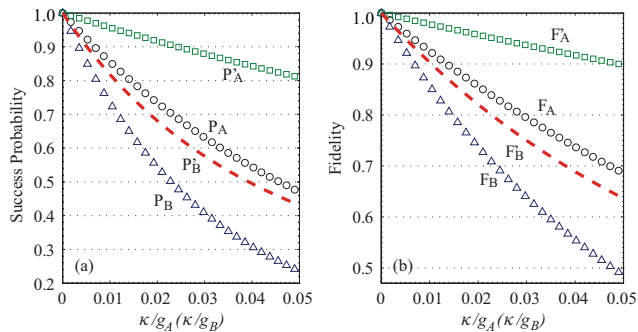}
\caption{(Color online) Contour plots for the success probability (a) and the fidelity (b) of the two-qubit (qutrit) entanglements in our scheme as a function of the cavity decay rate $\kappa$, where $P_A$, $P_B$ ($P'_A$, $P'_B$) denote the success probability of the two-qubit (qutrit) entanglements generated respectively in two stages, and  $F_A$, $F_B$ ($F'_A$, $F'_B$) denote the fidelity of the entanglements mentioned above.}
\label{fig:}
\end{figure}

In summary, we have proposed a scheme to prepare two kinds of atomic entangled states in a high-$Q$ bi-mode cavity. in the case of large atom-cavity detuning the scheme is immune to the effect of atomic spontaneous emission. Our computer numerical simulation results indicate that by choosing proper parameters we could entangle two atoms with muliti-dimensions in an optical cavity with high success probabilities and fidelities.

\begin{acknowledgments}
This work was supported by the National Natural Science Foundation of China (Grant No. 60677044, 11005099).
\end{acknowledgments}

\end{document}